\documentclass[twocolumn,amsmath,amssymb]{revtex4}
\usepackage{graphicx,dcolumn,bm}

\begin{document}

\title{Proposed precision laser spectrometer for trapped, highly charged ions}

\author{M. Vogel}
\email{m.vogel@imperial.ac.uk}
\author{D.F.A. Winters}%
\author{D.M. Segal}
\author{R.C. Thompson}

\affiliation{Blackett Laboratory, Imperial College London, Prince Consort Road, London SW7 2BW, United Kingdom}
\date{October 2005}

\begin{abstract}
We propose a novel type of precision laser spectrometer for trapped, highly charged ions nearly at rest. It consists of a cylindrical open-endcap Penning trap in which an externally produced bunch of highly charged ions can be confined and investigated by means of laser spectroscopy. The combination of confinement, cooling and compression of a dense ion cloud will allow the ground state hyperfine splitting in highly charged ions to be measured with an accuracy three orders of magnitude better than in any previous experiment. A systematic study of different charge states and different isotopes of the same element allows for highly sensitive tests of bound-state QED and for a precision determination of nuclear properties. Apart from stable isotopes, also radioactive species with half-lives longer than about one hour can be investigated.
\end{abstract}

\maketitle

\newpage

\section{Introduction}
The wavelength of the ground state hyperfine splitting (HFS) in atoms and ions is typically in the microwave domain and thus not accessible with lasers. However, the energy of the $1s$ ground state HFS of hydrogen-like ions scales with the atomic number $Z$ as $Z^3$ and shifts into the laser-accessible region above $Z \approx 60$ \cite{BEI00}. At the same time, the upper state lifetime scales as $Z^{-9}$ and eventually shifts into a region where acceptable fluorescence rates from magnetic dipole (M1) transitions can be expected. This allows for accurate measurements of the transition by means of laser spectroscopy and in turn for sensitive tests of corresponding calculations of the transition energy 
and lifetime \cite{SHA94,SHA97}. The $1s$ ground state HFS in hydrogen-like ions has until now been observed experimentally in only a few species: $^{165}\mbox{Ho}^{66+}$ \cite{KLA94}, $^{185,187}\mbox{Re}^{74+}$ \cite{CRE96}, 
$^{203,205}\mbox{Tl}^{80+}$ \cite{CRE98}, $^{207}\mbox{Pb}^{81+}$ \cite{SEE98} and $^{209}\mbox{Bi}^{82+}$ \cite{BEI01}. To our knowledge, two measurements of the $2s$ ground state HFS in lithium-like ions have been attempted. Both measurements were performed on $^{209}\mbox{Bi}^{80+}$ and the results are inconclusive \cite{BEI98,BOR00}.

The above measurements have been carried out at a strorage ring \cite{SEE98,BEI01} and a common experimental obstacle has been the effect of the Doppler width and shift of the transition due to the relativistic velocities of the ions. The measurements performed in an EBIT (electron beam ion trap) \cite{KLA94,CRE96,CRE98} are not as severely subject to this effect, but suffer from a low signal-to-noise ratio. The measurement scheme outlined here has the advantage that it takes place in cryogenic surroundings with well-localised particles nearly at rest. This reduces the Doppler width and shift to a level of 10$^{-7}$ of the measured transition, which is three orders of magnitude lower than in any previous experiment. Furthermore, by use of a \lq rotating wall\rq\;\cite{ITA98}, a high ion number density can be obtained which, together with the localisation, increases the intensity of the measured fluorescence.

This instrument is to be used within the framework of the HITRAP project \cite{QUI01} at the Gesellschaft f\"ur Schwerionenforschung (GSI) in Germany. The highly charged ions will be produced by the Experimental Storage Ring (ESR), subsequently slowed and cooled, and finally made available at low energies for experiments.

In section \ref{sectwo} we discuss the theoretical background to this area before describing the proposed instrument and measurement techniques in sections \ref{secthree} and \ref{secfour}.

\section{Hyperfine splitting calculations}\label{sectwo}
HFS calculations have been performed for a number of hydrogen-like ions \cite{BEI00,SHA94,SHA97}; the most accurate ones are given for the isotopes listed above. As a first approximation, good within 4 \%, the HFS of the $(1s)~^2$S$_{1/2}$ ground state of hydrogen-like ions is given by \cite{BEI00}:
\begin{equation} 
\label{one}
E_{HFS}=\frac{4}{3}\alpha (Z \alpha)^3 g_I \frac{m_e}{m_p} \frac{2I+1}{2} m_e c^2 A_{1s} (1-\delta_{1s} )
\end{equation}
with
\begin{equation} 
\label{two}
A_{1s}=\frac{1}{\kappa(2\kappa-1)} \quad \textrm{and} \quad \kappa=\sqrt{1-(Z\alpha)^2}
\end{equation}
where $\alpha$ is the fine structure constant, $g_I=\mu / (\mu_N I)$ is the nuclear $g$-factor (with $\mu$ the nuclear magnetic moment and $\mu_N$ the nuclear magneton), $I$  the nuclear angular momentum, $m_e$ and $m_p$ are the electron and proton mass respectively, and $c$ is the speed of light. Equation (\ref{one}) represents the normal ground state HFS multiplied by a correction $A_{1s}$ for the relativistic energy of the $1s$ electron, where $\kappa$ is related to the angular momentum $j$ of the electron. To account for the charge distribution of the nucleus, a first order approximation is made by assuming that the charge is evenly distributed over the volume of the spherical nucleus. This finite-size correction is given by the factor $(1-\delta_{1s})$ in equation (\ref{one}). The values for $\delta_{1s}$ have been taken from \cite{SHA94}, those for $g_I$ and $I$ were taken from \cite{FIR98}. The values for $^{207}\mbox{Pb}^{81+}$ for example, are $\delta_{1s}=0.1059$, $g_I=+0.59258$ and $I=1/2$. The corresponding M1 transition is between the (lower) $F=0$ and (upper) $F=1$ hyperfine states.

\begin{figure}[!t]
\begin{center}
\centering
\includegraphics[width=6cm]{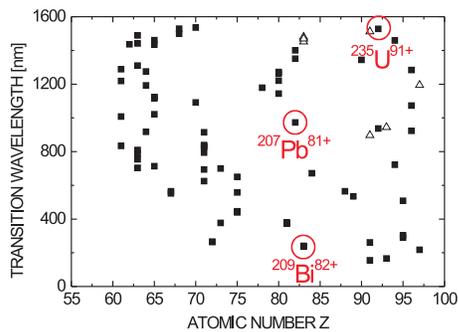}
\caption{Transition wavelength between the upper and lower hyperfine level of the ground state for different hydrogen-like (full symbols) and lithium-like ions (open symbols).}
\label{fig1}
\end{center}
\end{figure}

Figure \ref{fig1} shows the corresponding transition wavelengths between the upper and lower hyperfine level of the ground state for different hydrogen-like and lithium-like ions. Only isotopes with a lifetime longer than one hour are depicted. The wavelength region is restricted to the range accessible with readily available laser systems. The figure illustrates that a large number of isotopes can be studied.

The main uncertainty in the HFS calculations is caused by the Bohr-Weisskopf effect \cite{BEI00,BOH50}, which is due to the finite spatial distribution of the nuclear magnetisation. This limitation can be circumvented by a comparison of the ground state HFS in a hydrogen-like ion and its lithium-like counterpart which to first order rules out all nuclear effects and allows for bound-state QED effects to be isolated \cite{SHA01}. 

Correspondingly, a measurement of the HFS transition in a series of isotopes of the same hydrogen-like species rules out the effects of charge and isolates nuclear properties which can be probed with the single electron.

When the results from equation (\ref{one}) are compared with experimental values \cite{KLA94,CRE96,CRE98,SEE98,BEI01}, the difference ($\leq$ 4\%) is found to increase with $Z$ and equation (\ref{one}) generally underestimates the experimental values. This is mostly attributed to the Bohr-Weisskopf effect and to QED contributions (vacuum polarisation and self-energy) \cite{BEI00}. For example, in the case of $^{207}$Pb$^{81+}$ the measured wavelength is 1019.7(2)\,nm \cite{SEE98}, while a full calculation including all the above corrections gives 1020(4)\,nm \cite{BEI00}.

Similarly to equation (\ref{one}), the HFS of the $(1s^2 2s)~^2$S$_{1/2}$ ground state for lithium-like ions is given by
\begin{equation} 
\label{three}
E_{HFS}=\frac{1}{6}\alpha (Z \alpha)^3 g_I \frac{m_e}{m_p} \frac{2I+1}{2} m_e c^2 A_{2s} (1-\delta_{2s} )
\end{equation}
with the relativistic correction
\begin{equation} 
\label{four}
A_{2s}=2\frac{2(1+\kappa)+\sqrt{2(1+\kappa)}}{(1+\kappa)^2\kappa(4\kappa^2-1)}.
\end{equation}
The values for $^{209}\mbox{Bi}^{80+}$ for example, are $\delta_{2s}=0.1126$, $g_I=+0.41106$ and $I=9/2$. The corresponding M1 transition is between the (lower) $F=4$ and (upper) $F=5$ hyperfine states.

The results of these calculations are shown in figure \ref{fig1} as open symbols. The corrections to the values from equation (\ref{three}) are estimated to be about 4\% in this case as well. From x-ray measurements in $^{209}$Bi$^{80+}$ a tentative HFS transition wavelength of 1512\,nm was derived by subtraction of two measured x-ray transitions yielding the wavelength of the HFS-transition \cite{BEI98}. Calculations following this experiment predicted 1582\,nm \cite{TOM00}, 1564\,nm \cite{BOU00} and 1554\,nm \cite{SHA98}. A second direct measurement aimed for the latter value, but no resonance within the predicted limits was found \cite{BOR00}. A definitive measurement is therefore very desirable.

\section{Experimental apparatus and techniques}\label{secthree}
\subsection{Setup and procedure}
For these experiments a cylindrical open endcap Penning trap \cite{GAB89}, with an additional capture electrode at both ends, has been chosen. The ring electrode is axially split into four segments to allow for the rotating wall technique \cite{ITA98,GRU01} to be used.

\begin{figure}[!t]
\begin{center}
\centering
\includegraphics[width=8cm]{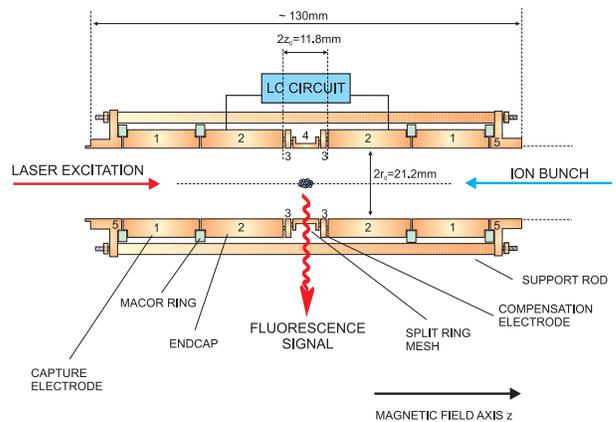}
\caption{Schematic of the trap setup enclosed in a cryogenic vacuum container.}
\label{fig2}
\end{center}
\end{figure}

The trap is to be enclosed in a vacuum container at liquid helium temperature, which ensures efficient cryopumping. The estimated residual gas pressure inside the trap is below 10$^{-12}$ Pa. This value is estimated from measured trapping times of highly-charged ions in a similar trap. Trapping is ensured by an electrostatic trapping potential across the trap and a static homogeneous axial magnetic field. As will be discussed below, it appears feasible to work with a magnetic flux density below 1\,T. For such a Penning trap experiment, typically a superconducting magnet would be used. However, fields as low as 1\,T could also be produced by permanent magnets, which are advantageous in terms of size and cost. Figure \ref{fig2} shows a schematic of the trap structure. Further details are given in \cite{WIN05}.

In the proposed experiment, an externally produced bunch of roughly $10^5$ ions at an energy of a few eV is loaded into the trap along the axis, {\it i.e.} along the magnetic field lines. It is captured in flight \cite{SCH86}, confined, resistively cooled and then radially compressed by the rotating wall technique. If necessary, several bunches of ions can be accumulated in the trap \cite{HAS94}. The spectroscopy laser driving the hyperfine transition is run in a continuous-wave mode and enters the trap axially through an open-endcap. The laser beam is shaped such that it 
illuminates the full radial cross section of the ion cloud. The fluorescence is detected perpendicular to the cooled axial motion (trap axis).

\subsection{Resistive cooling}
The axial motion is resistively cooled by a resonant circuit attached to the endcaps. The ion motion induces image charges in the endcaps and causes an oscillating current to flow through the resonant circuit, which is tuned to the axial trapping frequency and effectively dissipates the ions' kinetic energy. The resistance of the circuit thus leads to a damping of the ion motion amplitude. The induced current is proportional to the ion velocity and therefore the dissipated power is proportional to the kinetic energy. For a single particle there is an exponential energy decrease, which is given by \cite{DEH68}
\begin{equation} 
\label{five}
E(t)=E_0 \mbox{exp} (-t/\tau) \quad \textrm{and} \quad \tau=\frac{m (2z_0)^2}{R q^2}.
\end{equation}
Here, $q$ is the particle charge, $m$ its mass and $2z_0$ is the effective distance between the endcaps \cite{GAB89}. In resonance, the impedance of an LC-circuit is real and acts as an ohmic resistor with resistance $R=QL\omega_z$, where $Q$ is the quality factor of the circuit and $L$ the inductance. The axial frequency $\omega_z$ is given by
\begin{equation} 
\label{omegaz}
\omega_z = \sqrt{\frac{qU_0}{mz_0^2}}
\end{equation}
where $U_0$ is the trapping potential. For cryogenic LC-circuits typical values are $Q \approx 500$, $L \approx 1$\,mH, and $C < 100$\,pF, so that at axial frequencies around 1\,MHz the resonance resistance is of the order of 1\,M$\Omega$. The corresponding bandwidth of the filter, defined as $\Delta \omega=\omega / Q$, is about 2\,kHz. The resistive cooling time constant, expressed in terms of the filter components, is given by
\begin{equation} 
\label{tau}
\tau=\frac{4(z_0 \sqrt{m})^3}{LQ\sqrt{q^5U_0}}.
\end{equation}
Deviations from the exponential cooling behaviour may occur when the dissipated power is not proportional to the kinetic energy of the ions. For example, if the axial frequency $\omega_z$ depends on the ions' kinetic energy, $\omega_z$ may move out of resonance with the LC-circuit. For large amplitudes the motion is no longer harmonic since higher order multipole components in the trapping potential may become significant \cite{MAJ04,BRO86}. The quality factor $Q$ of the resonant circuit thus needs to be chosen such that both cooling time and bandwidth are optimal.

\begin{figure}[!t]
\begin{center}
\centering
\includegraphics[width=6cm]{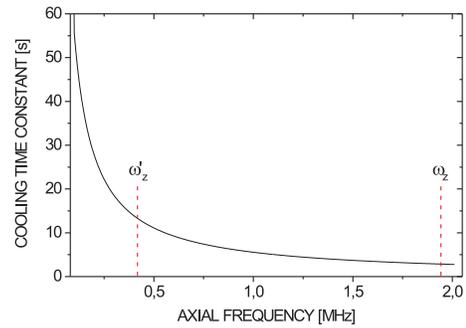}
\caption{Calculated resistive cooling time constant of Pb$^{81+}$ ions as a function of the axial frequency. Also shown is the shift due to space charge effects (section \ref{spce}). It is assumed that the resonant circuit stays tuned to the axial frequency.}
\label{fig3}
\end{center}
\end{figure}

In a cloud of ions, only the common centre-of-mass-mode (c.m.-mode) is cooled resistively with high efficiency \cite{WIN75}. This is easily understood for the so-called \lq breathing mode\rq, where the centre of the charges is 
stationary and therefore (to first order) no current is induced. However, through intra-cloud interactions, such as ion-ion collisions, energy transfer between the ions occurs thus creating motional fluctuations which induce currents leading to energy dissipation from the motion.

The final temperature is determined by electronic noise present in the resonant circuit, which effectively heats the ion cloud \cite{DJE04}. The electronic noise temperature can significantly exceed the ambient temperature \cite{DJE04,FIL95}, and we therefore assume an axial ion temperature of 10\,K, instead of 4.2\,K.

The c.m.-motion of an ion cloud obeys the single particle equation (\ref{five}). To first order $\tau$ does not depend on the number of ions under the condition that the distribution of axial frequencies is smaller than the bandwidth of the resonant circuit \cite{MAJ04}.

Figure \ref{fig3} shows the dependence of $\tau$ on $\omega_z$ for $^{207}$Pb$^{81+}$-ions cooled by a resonant circuit ($Q$=500, $L$=1\,mH) attached to the trap ($z_0$=15\,mm, $U_0$=1000\,V). The figure assumes that the resonant circuit is always tuned to the axial frequency. A lowering of $\omega_z$, for example due to space charge effects (section \ref{spce}), results in an increase of $\tau$.

\begin{figure}[!t]
\begin{center}
\centering
\includegraphics[width=8cm]{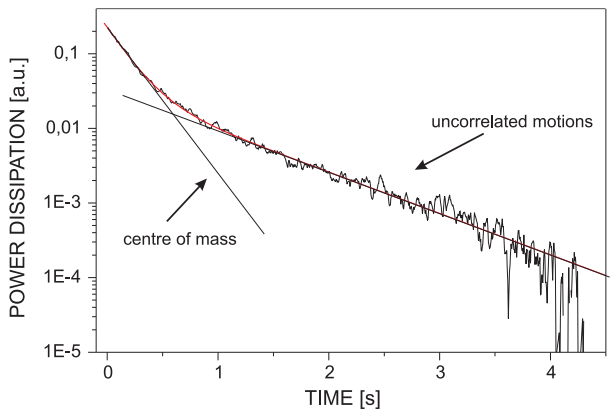}
\caption{Resistive cooling behaviour of a cloud of 30 C$^{5+}$-ions, taken from \cite{HAF03}.}
\label{fig4}
\end{center}
\end{figure}

Figure \ref{fig4} shows experimental data on the cooling of a cloud of about 30 $^{12}$C$^{5+}$-ions with an initial mean kinetic energy of about 13\,eV in a potential well of 50\,eV (taken from \cite{HAF03}). Due to the high energy, trapping potential anharmonicities shift the axial frequency downwards by about 5\% so that only the coldest fraction of the ions in this ensemble is in resonance with the cooling circuit. Within the Boltzmann distribution of ion energies, this is the largest fraction and it results in the initial fast decrease of the total noise power in figure \ref{fig4}. The cooling time constant (around 132\,ms) is equal to the one measured for an ion cloud with low kinetic energy in a harmonic trap. On longer timescales, the dominant cooling process is the energy dissipation from the non-c.m.-modes. This has a measured value of $\tau=5$\,s, which is more than one order of magnitude longer than for the c.m.-mode. For the planned experiments, we expect $\tau$-values of several tens of seconds.

\subsection{Rotating wall}
When an ion plasma confined in a Penning trap is coupled such that the dynamics cannot be described by a single particle approach, the ion number density $n$ (see eq. \ref{dens}) and related properties of the cloud are governed by its global rotation frequency $\omega$. This is the case if the Debye length 
\begin{equation} 
\lambda_D = \sqrt{\frac{\epsilon_0k_BT}{2nq^2}}
\end{equation}
is much smaller than any of the cloud dimensions. The density drops from its homogeneous value to zero over a 
distance $\lambda_D$, but as this distance is much smaller than the cloud dimensions, it can be treated as a hard edge. The plasma acts as a solid when the coupling parameter $\Gamma$ defined as
\begin{equation} 
\Gamma = \frac{q^2}{4\pi\epsilon_0ak_BT}
\end{equation}
is sufficiently large. Here, $a$ is the Wigner Seitz radius defined by $4 \pi n a^3/3 = 1$, $\epsilon_0$ is the permittivity of free space, $k_B$ is Boltzmann's constant and $T$ the ion ensemble temperature. If $\Gamma \ll 1$ the system of ions forms a weakly correlated plasma. For $\Gamma \geq 2$ the cloud shows fluid characteristics. Above $\Gamma \approx 170$ the ions form a rigid crystal lattice with homogeneous density \cite{BOL95}. 

For the cloud of 30 C$^{5+}$ ions, whose cooling behaviour is shown in figure \ref{fig4}, the estimated Debye length is comparable to the mean particle spacing of a few tens of micrometres. The estimated coupling parameter is about 5 and the ion cloud behaves like a fluid. 

For a plasma formed of highly-charged ions, like $^{207}$Pb$^{81+}$ at cryogenic temperatures and with a density of some 10$^6$\,cm$^{-3}$, the Debye length is several micrometres, which is much smaller than typical cloud dimensions of millimetres. At the same time, $\Gamma$-values of 300 can easily be reached.

The cloud rotation is driven by the electric field produced by a rotating dipole. To this end, sinusoidal AC-voltages with a fixed phase shift are applied to the different ring segments of the trap \cite{ITA98}. The rotating wall technique was first applied to increase the cloud density of a laser cooled ion cloud in order to observe crystallisation \cite{ITA98,GRU01}. The resulting cloud compression is due to the Lorentz force directed towards the trap centre. The final density results from a balance between the Lorentz force and the Coulomb force between the ions. 

It can be shown \cite{DUB99} that the ion number density of a single-component plasma, such as a cloud of highly charged ions, is given by
\begin{equation} 
\label{dens}
n=\frac{2 \epsilon_0 m \omega (\omega_c-\omega)}{q^2},
\end{equation}
where $\omega$ is the applied rotating wall frequency.
The single-ion cyclotron frequency is given by $\omega_c = qB/m$, where $B$ is the magnetic flux density. Possible values for $\omega$ range from $\omega_m$ to $\omega'_{c}$, which are the single-ion magnetron frequency
\begin{equation} 
\omega_m = \frac{\omega_c-\sqrt{\omega_c^2-2\omega_z^2}}{2}
\end{equation}
and the reduced cyclotron frequency
\begin{equation} 
\omega'_{c} = \frac{\omega_c+\sqrt{\omega_c^2-2\omega_z^2}}{2}
\end{equation}
The minimum density occurs when $\omega=\omega_m$ or $\omega=\omega'_c$:
\begin{equation} 
\label{n_min}
n_{min} = \frac{\epsilon_0 U_0}{qz_0^2}
\end{equation}
and the maximum when $\omega = \omega_c /2$ (\lq Brillouin limit\rq):
\begin{equation} 
n_{max}=\frac{\epsilon_0 m \omega_c^2}{2q^2}=\frac{\epsilon_0 B^2}{2m}
\end{equation}
Rotating wall frequencies below the actual cloud rotation frequency lower $n$ and vice versa \cite{ONE98,DUB99}.

Although $n$ is solely determined by $\omega$, the shape of the ion cloud depends on both $\omega$ and $U_0$ and is generally a spheroid. Once the aspect ratio $\beta$ (length divided by diameter) is determined, the dimensions of the ion cloud can be calculated, provided that the density $n$ and the number of particles $N$ are known. The aspect ratio is implicitly related to $\omega_z$, and therefore to $U_0$, by a transfer function $b(\beta)$ defined by \cite{BRE88}
\begin{eqnarray} 
\label{betaeq}
b(\beta) &=& \frac{3}{\beta^2-1}\,Q^0_1\left(\frac{\beta}{\sqrt{\beta^2-1}}\right), \quad b(\beta)=1, \\ \nonumber
b(\beta) &=& \frac{3}{\beta^2-1}\left[\frac{\beta}{\sqrt{1-\beta^2}} \arctan\left(\frac{\sqrt{1-\beta^2}}{\beta}\right)-1\right]
\end{eqnarray}
for $\beta>1$, $\beta=1$ and  $\beta<1$, respectively. $Q^0_1$ is the associated Legendre function of the second kind
\begin{equation} 
Q^0_1 (u) = \frac{u}{2} \ln \left(\frac{u+1}{u-1}\right) -1.
\end{equation}
The function $b(\beta)$ has the following relation with $\omega_z$ or $U_0$:
\begin{equation} 
b(\beta) = \frac{3\omega_z^2}{2\omega(\omega_c-\omega)} = \frac{3qU_0}{2m z_0^2\omega(\omega_c-\omega)}
\end{equation}
It is instructive to see what values $b(\beta)$ takes for the three limits of the frequency, {\it i.e.} for $\omega_m$, $\omega'_{c}$ and $\omega_c/2$:
\begin{eqnarray} 
b(\beta_{min}) &=& \frac{3\omega_z^2}{2\omega_m(\omega_c-\omega_m)} = \frac{3\omega_z^2}{2\omega'_{c}(\omega_c-\omega'_{c})}=3 \\ \nonumber
b(\beta_{max}) &=& \frac{3\omega_z^2}{2\omega_c/2(\omega_c-\omega_c/2)} = \frac{6mU_0}{z_0^2qB^2}
\end{eqnarray}

This means that $b(\beta)=3$ when $\omega$ equals $\omega_m$ or $\omega'_{c}$, irrespective of the particle and trapping parameters. At these frequency limits $\beta$ vanishes, {\it i.e.} the cloud looks like an inifinitely large and thin disc (oblate). For any other frequency $b(\beta) < 3$, so that $\beta > 0$. (Note: equations (\ref{betaeq}) cannot be analytically solved for $\beta$.)

Therefore, the cyclotron and rotating wall frequencies determine the ion number density of the cloud, the number of ions changes the cloud dimensions, and the trapping potential determines the aspect ratio. The lower $U_0$, the more elongated (prolate) the cloud is. For fixed $U_0$ and $B$, both shape and dimensions are fully determined by $\omega$. For the present experiment, these parameters are chosen to produce a nearly spherical ion cloud, this is the case for example for $B$=0.8\,T, $U_0$=1000\,V and $\omega=\omega_c/2$ when about $10^5$ ions are confined in the trap.

\subsection{Related effects}
\subsubsection{Exerted torque}
When the rotating wall frequency $\omega$ differs from the actual rotation frequency of the ion cloud, the torque exerted is not maximal and \lq slip\rq\;occurs. This leads to a cloud rotation frequency lower than $\omega$, even if no internal degrees of freedom of the plasma are heated (see below). Therefore, $\omega$ is ramped up from $\omega_m$ to the desired rotation frequency. By this method densities are increased in a controllable manner, as has been shown in \cite{HUA97,HUA98b,HUA98,AND98,HOL00}. The ideal ramping speed is a compromise between slip and undesired excitation of plasma modes. The exerted torque $\xi$ has been empirically found to follow qualitatively the following relationship \cite{MAJ04}:
\begin{equation} 
\xi \propto  \frac{1}{|\omega-\omega'|} \frac{A_d}{\sqrt{T}}
\end{equation}
where $A_d$ is the amplitude of the drive, $T$ is the plasma temperature and $|\omega-\omega'|$ is the difference between the applied and the actual rotation frequency of the cloud.

\subsubsection{Excitation of plasma modes}
Usually it is found that at a certain driving frequency the ion cloud density stabilises and then drops to zero for higher frequencies. This has been found to occur due to plasma resonances. The (2,1) plasma mode was the limiting factor to the density in the experiments by Bollinger et al \cite{BOL93}. The (2,1) mode is a precession motion of the plasma's main axis about the magnetic field lines, as is depicted for example in \cite{BOL95,BOL93}. If the applied frequency matches such a plasma resonance, energy is transferred from the external field into deformation motions of the plasma rather than into global rotation \cite{BRE88,BOL93,HEI91,TIN96,DUB96}. Due to this energy transfer the cloud heats up and eventually ions will be lost \cite{DUB91,GOS98,GOS00}.

\begin{figure}[!t]
\begin{center}
\centering
\includegraphics[width=6cm]{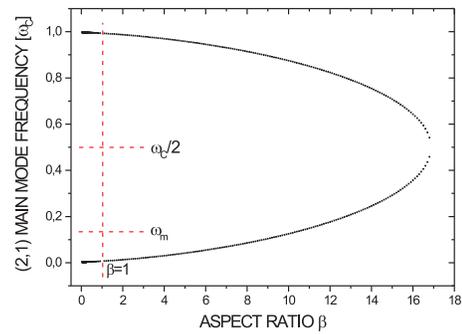}
\caption{(2,1)-plasma mode frequency ($k=1$) versus aspect ratio $\beta$. For $\beta \approx 1$ there is no (2,1) plasma mode frequency between $\omega_m$ (minimum) and $\omega_c/2$ (maximum).}
\label{fig6}
\end{center}
\end{figure}

The mode spectrum depends on the applied trapping potentials and by appropriate sudden switching, a plasma mode may jump over the applied frequency \cite{BOL93}. To be able to make use of this, the plasma mode frequencies need to be known to sufficient accuracy. For the (2,1) plasma mode the three mode frequencies $\omega_1$, $\omega_2$ and $\omega_3$ result from the three solutions $\eta_1$, $\eta_2$ and $\eta_3$ of the cubic 
equation $a_3\eta^3+a_2\eta^2+a_1\eta+a_0=0$ \cite{BOL93}:
\begin{equation} 
\omega_k =(-1)^{k+1} \sqrt{-\omega_p^2 f_{\beta} \frac{\beta^2/\eta_k+1}{3}},
\end{equation}
for $k=1,2,3$. Here $\omega_p$ and $f_{\beta}$ are defined as:
\begin{equation} 
\omega_p=\sqrt{\frac{q^2n}{\epsilon_0m}} \quad \textrm{and} \quad 
f_{\beta} = \frac{3}{2} \left( \frac{1-3 \omega_z^2/\omega_p^2}{1-\beta^2} \right)
\end{equation}
where $\omega_p$ is the plasma frequency. The coefficients $a_3$, $a_2$, $a_1$ and $a_0$ are
\begin{eqnarray} 
a_3 &=& \frac{1}{3}\omega_p^2 f_{\beta}, \quad a_2=a(\beta^2-2) \\ \nonumber 
a_1 &=& a(1-2\beta^2)+\omega_v^2, \quad a_0=a\beta^2
\end{eqnarray}
where $\omega_v$ is the plasma vortex frequency given by $\omega_v=\omega_c-2\omega'_{c}$. In figure \ref{fig6} the plasma mode 
frequency (in units of $\omega_c$) for $k=1$ is plotted as a function of the aspect ratio $\beta$ for $^{207}$Pb$^{81+}$ ions at $T$=10\,K and $B$=0.8\,T. The frequency for $k=2$ has opposite sign and the frequency for $k=3$ is negligibly small. For the envisaged aspect ratio of $\beta \approx 1$ there is no (2,1) plasma mode between the minimum ($\omega_m$) and maximum ($\omega_c/2$) rotating wall frequencies.

\subsubsection{Space charge effects}
\label{spce}
Due to the high density of highly charged ions in the trap, space charge effects occur, i.e. the positive potential due to the ions effectively reduces the quadrupole potential. To compensate for this effect, the trapping potential $U_0$ needs to be increased. The most prominent space charge effect is a large downward shift of the axial frequency $\omega_z$, as depicted in figure \ref{fig3}. Image charges induced in the trap electrodes also lead to frequency shifts but these are negligible compared to the space charge shifts.

\begin{figure}[!t]
\begin{center}
\centering
\includegraphics[width=8cm]{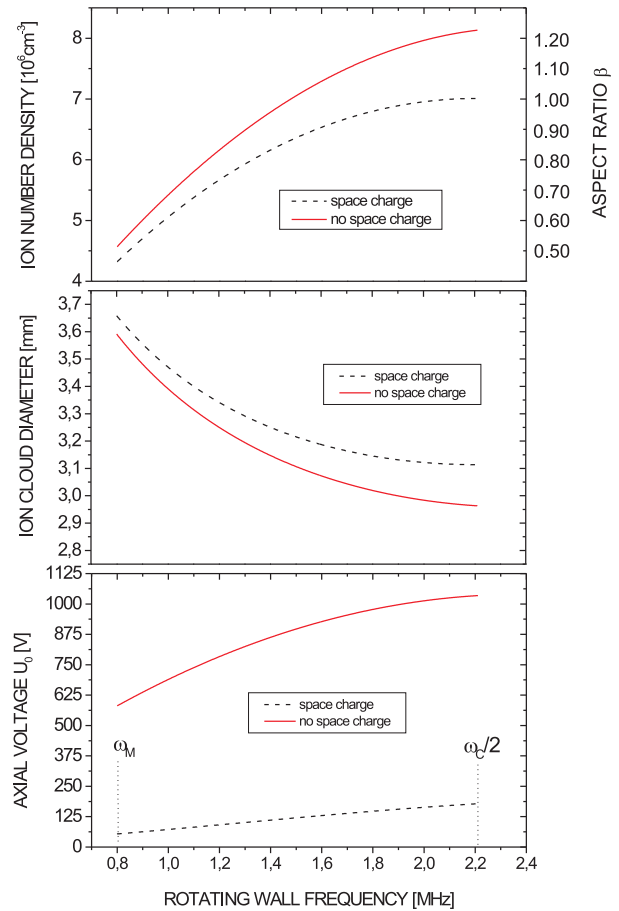}
\caption{Ion cloud properties with and without space charge effects plotted as a function of the rotating wall frequency: (top) ion number density $n$ and aspect ratio $\beta$, (middle) diameter of a spherical cloud ($\beta \approx 1$), (bottom) trapping potential.}
\label{fig7}
\end{center}
\end{figure}

When the density of the spherical ($\beta \approx 1$) ion cloud is nearly homogeneous, the shifted axial frequency for a fixed trapping potential is \cite{MAJ04}
\begin{equation} 
\omega'_z=\omega_z \left( 1- \frac{\omega_p^2}{3\omega_z^2} \right)^{1/2}
\end{equation}
The space charge effects thus lower $\omega_z$, increase the resistive cooling time constant $\tau$, and lower the plasma density $n$. Figure \ref{fig7} shows the space charge effects on the ion number density, the cloud dimension and the trapping potential as a function of the rotating wall frequency $\omega$. Parameters used are as before, $T$=10\,K, $B$=0.8\,T, $N$=10$^5$ $^{207}$Pb$^{81+}$ ions, $z_0$=15\,mm. Note that the space charge effects reduce the effective trapping potential by almost one order of magnitude.

\begin{figure}[!t]
\begin{center}
\centering
\includegraphics[width=8cm]{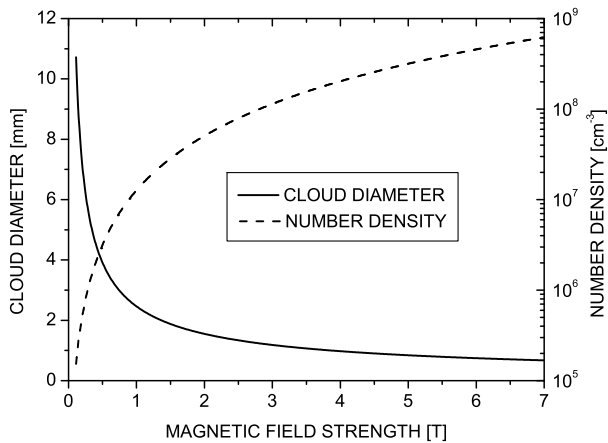}
\caption{Calculated cloud diameter for a spherical cloud ($\beta \approx 1$) and maximum ion number density $n$ plotted as a function of the magnetic flux density (space charge effects included).}
\label{fig0}
\end{center}
\end{figure}

Figure \ref{fig0} shows the calculated cloud diameter for a spherical cloud ($\beta \approx 1$) and the maximum achievable ion number density $n$ as a function of the magnetic flux density $B$, including space charge effects. While the ion number density increases strongly with increasing magnetic flux density, even low magnetic fields already lead to acceptable cloud sizes. The cloud density is important for the fluorescence rate, as will be discussed in the next section.

\section{Laser spectroscopy}\label{secfour}
A possible pitfall in the laser spectroscopy experiment proposed here is the danger of optical pumping into undesired states. When the level scheme is such that the population may be pumped to hyperfine substates ($M_F$) that cannot be excited by the pumping laser, these ions remain in the trap but no longer contribute to the fluorescence. However, as long as the degeneracy of the upper hyperfine level exceeds that of the lower one, and circularly polarised light is used, optical pumping is not a problem. This is the case if $g_I$ is positive, which is true for example for $^{207}$Pb$^{81+}$ \cite{SEE98} and $^{209}$Bi$^{80+}$ \cite{BEI98}.

The transition probability for a magnetic dipole (M1) transition from the excited to the lower hyperfine state is given by \cite{BEI00}:
\begin{equation} 
A=\frac{4 \alpha \omega^3 \hbar^2 I \left( 2\kappa+1 \right)^2}{27m_e^2 c^4 \left( 2I+1 \right)}
\end{equation}
$\alpha$, $\kappa$, and $I$ were defined in the discussion of equation (\ref{one}). The lifetime of the excited state is $t=A^{-1}$, and the natural linewidth of the transition is $\Delta \nu=(2 \pi t)^{-1}$. The relative Doppler-broadened linewidth of the transition is given by \cite{DEM96}
\begin{equation} 
\frac{\Delta \nu_D}{\nu}=\frac{2 \sqrt{2 ln 2}}{c} \sqrt{\frac{k_B T}{m}}.
\end{equation}

The saturation intensity $I_s$ of a hyperfine transition with wavelength $\lambda$ is given by \cite{DEM96}:
\begin{equation} 
I_s=\frac{2 \sqrt{2} h \nu A}{\lambda^2}
\end{equation}
For most species shown in figure \ref{fig1}, saturation intensities less than 1000\,Wm$^{-2}$ are required, which are readily obtained in the relevant wavelength region by use of common laser systems. The values for $^{207}$Pb$^{81+}$, $^{209}$Bi$^{80+}$ and $^{235}$U$^{91+}$ are 153\,Wm$^{-2}$, 71\,Wm$^{-2}$ and 28\,Wm$^{-2}$, respectively. The radial cross section of the ion cloud is of the order of 10\,mm$^2$, which implies that a maximum laser power of only a few mW is sufficient.

The relative Doppler-broadened linewidth $\Delta \nu_D/\nu$ of the transition slowly decreases almost linearly with atomic number $Z$ for hydrogen-like and lithium-like ions. For hydrogen-like lead $^{207}$Pb$^{81+}$, $\Delta \nu_D / \nu$ is about $1.6 \times 10^{-7}$, which corresponds to a Doppler broadening of 50\,MHz at 10\,K, while $\Delta \nu=3$\,Hz. The anticipated accuracy of a measurement on cold, trapped ions is therefore of the order of $10^{-7}$, which is three orders of magnitude better than any previous measurement \cite{KLA94,CRE96,CRE98,SEE98,BEI01}.

For an upper state lifetime of about 50\,ms in the case of $^{207}$Pb$^{81+}$, and for a detection efficiency of the order of some $10^{-3}$, the expected fluorescence rate from a fully saturated cloud of $10^{5}$ ions is a few thousand counts per second. As compared to expected background rates below 100\,s$^{-1}$, this yields a sufficient signal-to-noise ratio of about 50.

\section{Conclusion}
We present a novel precision laser spectrometer for measurements on cold, trapped highly charged ions. An externally produced bunch of highly charged ions from a storage ring is captured in flight, confined in a cylindrical open-endcap Penning trap, resistively cooled to cryogenic temperatures and radially compressed by a rotating wall. Subsequently, it is irradiated by a laser and fluorescence light is detected perpendicular to the trap axis. A measurement of the ground state hyperfine splitting in a highly charged ion represents a sensitive test of corresponding calculations, which include the Bohr-Weisskopf effect and QED effects. By a comparison of the ground state HFS in different charge states of the same isotope all nuclear effects can be ruled out to first order, which allows for bound-state QED effects to be isolated. Suitable species are for example $^{207}$Pb$^{81+}$, $^{209}$Bi$^{82+}$, $^{209}$Bi$^{80+}$ and $^{235}$U$^{91+}$. With a sufficiently dense and well-localised cloud of ions, a laser spectroscopy experiment with a relative accuracy of a few parts in $10^{7}$ is feasible. This accuracy exceeds that of previous measurements by three orders of magnitude. Essential prerequisites are sufficient cooling of the ion motion by resistive cooling, and ion cloud compression by use of a rotating wall technique. Space charge effects have been included in the calculations to account for the influences of the high charge density within the ion cloud. The presented methods are applicable also to radioactive species with half-lives longer than the typical measurement time of the order of 1 h.

\begin{acknowledgments}
This work is supported by the European Commission within the framework of the HITRAP project (HPRI-CT-2001-50036).
\end{acknowledgments}

\end{document}